# Unravelling the optomechanical nature of plasmonic trapping

Optomechanical Plasmonic Trapping

Pau Mestres[1], Johann-Berthelot[2] and Romain Quidant[3,*]

[1] *ICFO- Institut de Ciencies Fotoniques, The Barcelona Institute of Science and Technology, 08860 Castelldefels (Barcelona), Spain*

*E-mail: pau.mestres@icfo.es*

[2] *ICFO- Institut de Ciencies Fotoniques, The Barcelona Institute of Science and Technology, 08860 Castelldefels (Barcelona), Spain*

*Institut FRESNEL - Faculte des Sciences de Saint Jérôme, 13397 Marseille Cedex, France*

*E-mail: johann.berthelot@fresnel.fr*

[3] *ICFO - Institut de Ciencies Fotoniques, The Barcelona Institute of Science and Technology, 08860 Castelldefels (Barcelona), Spain*

*ICREA - Institució Catalana de Recerca i Estudis Avançats, 08010 Barcelona, Spain*

*\*Corresponding author E-mail: romain.quidant@icfo.es*
*Corresponding author phone number: +34 935534076*

# Abstract

Non-invasive and ultra-accurate optical manipulation of nanometer objects has recently gained a growing interest as a powerful enabling tool in nanotechnology and biophysics. In this context, Self-Induced Back-Action (SIBA) trapping in nano-optical cavities has shown a unique potential for trapping and manipulating nanometer-sized objects under low optical intensities. Yet, the existence of the SIBA effect has that far only been evidenced indirectly through its enhanced trapping performances. In this article we present for the first time a direct experimental evidence of the self-reconfiguration of the optical potential experienced by a nanoparticle trapped in a plasmonic nanocavity. Our observations enable us gaining further understanding of the SIBA mechanism and determine the optimum conditions to boost the performances of SIBA-based nano-optical tweezers.

# Introduction

Optical trapping and manipulation have emerged as powerful tools to investigate single microscopic objects in a controlled environment. Using the momentum carried by light, forces can be exerted to confine and manipulate objects in a wide range of conditions ranging from ultra-high pressures to high vacuum [1-4]. Trapped objects experience two types of optical forces: a scattering force pushing the object along the direction of propagation of light, and a gradient force pulling the object towards the maximum of intensity [5]. In order to form a stable potential, gradient forces need to overcome the scattering contribution in all three dimensions. In conventional optical tweezers, this can be achieved by tightly focusing a laser beam with a high NA objective [6, 7], creating a large intensity gradient in a diffraction limited spot. For displacements smaller than the wavelength of light, the trapped particle experiences a linear restoring force towards the centre of the trap.

As the object decreases in size so does its polarizability, thus making it more challenging to obtain a stable potential that overcomes environmental fluctuations. Consequently, trapping nanoscale objects with conventional optical tweezers, can require few hundreds of mW of optical power focused onto a diffraction limited spot size. Although some dielectric objects can withstand trapping at such high optical intensities [8-10], this is not generally the case. For example, local absorption in sensitive samples is known to induce conformational changes in proteins [11], arrest of the cell cycle [12], photobleaching and promotion of dark states in NV centres [13] among others.

To circumvent these difficulties, researchers proposed using rapidly decaying evanescent fields from plasmonic nanostructures to create larger field gradients over distances $L \ll \lambda$ [14, 15]. Although first experimental implementations of plasmon-assisted trapping [16-19] showed great potential, they remained limited, because of photothermal effects, to object sizes greater than 100nm.

To further improve the trapping efficiency, an alternative strategy inspired from optomechanics can be adopted. In this strategy the trapped specimen plays an active role in the trapping mechanism, the so called self-induced back-action (SIBA) effect, thus requiring much lower local intensities as compared to a static potential [20]. Following this approach, dielectric objects of tenths of nm size and individual biomolecules have been trapped with less than 10 mW of optical power [21, 22].

In these experiments, plasmonic nanocavities are milled in an opaque metal layer. The sensitivity of the plasmon resonance to local changes in the refractive index [25] is at the origin of the SIBA effect and also enables to monitor the trapping dynamics through the changes in transmitted light [24]. The SIBA hypothesis has been indirectly validated by the enhanced trapping performances observed in the experiments [20, 21, 26]. However, both direct observation of the optical potential modulation and the optimum conditions for SIBA remain to be demonstrated. Further understanding of this effect and the regimes under which it takes place would pave the way to improved performance of nanoscale trapping and manipulation of tiny objects with low laser intensities.

In this article we evidence the optomechanical nature of the back-action effects involved in a plasmonic nanotrap. By enhancing the nanoparticle/nanocavity coupling we are able to directly monitor the trapping potential modulations. Finally a study of the different cavity detuning regimes allows us to identify the optimum conditions for SIBA trapping, which correspond to a previously unstudied regime.

## Materials and Methods

In our experiment, we consider a nanosphere trapped in a nanocavity surrounded by liquid. For small displacements from the centre of the trap ($|x| \ll \lambda$), the trapped object follows the overdamped Langevin equation of motion:

$$\gamma \dot{x} + \kappa_{tot} x = \xi(t),$$

where $\gamma$ is the viscous damping, $\kappa_{tot}$ the stiffness of the trap and $\xi(t)$ the thermal fluctuations. Since the particle radius $a$ is much smaller than the wavelength of light ($a \ll \lambda$), the optical forces experienced by the particle are given by [27]:

$$\vec{F}_{grad} = \frac{\alpha}{4} \nabla I_o(r),$$

where $I_o(r)$ is the optical intensity profile and $\alpha$ is the polarizability of the object, that scales with its volume $V$ ($\alpha \propto V$). Due to the dispersive coupling between the cavity and the object, the former induces a frequency shift [28]:

$$\delta\omega_o(r_p) = \omega_c \frac{\alpha}{2V_m\epsilon_o} f(r_p),$$

where $\omega_c$ is the optical resonance frequency, $V_m$ the mode volume and $f(r)$ the cavity intensity profile. By this mechanism, changes in the particle position modify the number of photons coupled into the cavity, and thus the optical potential. In the case of an adiabatic cavity response, $\kappa_{tot}$ can be decomposed into sum of two contributions [29]:

$$\kappa_{tot} = \kappa_{opt} + \kappa_{SIBA}$$

with $\kappa_{opt}$ depending on the system (cavity + trapped particle) resonance profile and $\kappa_{SIBA}$ originating from changes in the particle position affecting the intra-cavity field. To optimize $\kappa_{SIBA}$, the value of $\delta\omega_o(r_p)$ with the cavity linewidth $\Gamma$, defined as the back-action parameter needs to be maximized [29]:

$$\upsilon = \frac{\delta\omega_o(r_p)}{\Gamma}$$

Although plasmonic cavities can feature very small mode volumes well below the diffraction limit [30], they suffer from large losses caused by the intrinsic absorption of metals [31], resulting in a fast cavity response and broad plasmonic resonances. As a consequence, $\Gamma$ and $V_m$ have limited tunability and are determined by the geometry and the fabrication process. In our experiment, the plasmonic nanocavity consists of bowtie nano-apertures (BNA) milled by focused ion beam (FIB) in a 100 nm thick Au film. Once patterned, the BNA are inserted in a liquid chamber containing a dilute suspension of gold nanoparticles (GNP) (BBI solutions) in trimethylammoniun bromide (MTAB) at 10 mM. To boost the optomechanical interaction, we need to allocate relatively large spheres that maximize the polarizability to mode volume ratio while maintaining the system resonance close to a trapping laser line. Considering these constraints, we chose an 85 nm gap BNA to trap 60 nm diameter gold nanospheres with an excitation laser at 1064 nm. From our previous finite element simulations of these antennae [21], we estimate a frequency shift $\delta\omega_o(r_p) \approx 125$nm that corresponds to an optomechanical coupling constant $G \equiv \frac{\delta\omega}{\delta x} \approx 750$GHz/nm. Note that due to the exponential decay of the near-field, the frequency shift mainly occurs for very small displacements (few nm) leading to an underestimation of the actual value of $G$. Still, this value compares well with the one reported in a previous plasmonic optomechanical system with similar dimensions [32] and is much greater than the typical values attained with standard optomechanical systems [33].

The sample was mounted upside-down on a home-made inverted microscope [20]. A continuous-wave 1064 nm Nd-YAG laser beam was focused onto the sample with a 40x microscope objective (0.65 NA). This low NA that corresponds to a spot size of about 2 µm diameter avoids direct laser trapping. Using two polarizers and a half-wave plate we control the polarization and the power of the incident beam, which is limited to a maximum of 10mW at the sample plane. Finally, the transmission of the trapping laser through the nanocavity is collected with a 20x NIR objective (0.40 NA) and sent to an avalanche photodiode (APD) (Figure 1a). The APD signal is recorded at 1MHz with a high resolution (12-bits) digital oscilloscope (Keysight S-Series). Simultaneously, we monitor the trapping events by splitting the APD signal to a 1 kHz sampling rate DAQ card. To tune the resonance conditions of our system, we fabricated an array of increasing size BNA in which the aspect ratio was kept constant as well and the central gap along the $x$ axis (85nm) (Figure 1b). We monitored the optical transmission with the incident laser polarized along $x$ (Figure 1b). As expected, the transmission increases with the length size of the antenna until optimum resonance is reached (antenna 11), and then it decreases.

When a GNP gets trapped in the nanocavity, the red-shift $\delta\omega_o(r_p)$ leads to one of the three different situations depicted in Figure 1c-d. We refer to these different regimes as: blue-shifted, resonant and red-shifted (Figure 1c (i), (ii) and (iii)) respectively. In the blue-shifted regime, the cavity mode is set well blue-detuned from the excitation wavelength. As soon as an object is trapped, the resonance red-shifts towards the laser line, increasing the local field and transmitted light (Figure 1c (i)). This case is the one most reported in the literature [21-23, 34]. Conversely, in the red-shifted regime, the presence of the particle leads to a transmission decrease (Figure 1c (iii)). In this regime, trapping is highly inefficient due to the strong negative optical response of the system in the presence of the particle. Finally, in the resonant regime, the cavity mode is set to be slightly blue shifted from the excitation laser. When trapping occurs, the system symmetrically red-shifts through the resonance, resulting in the transmissions of *empty* and *trapping* states to be comparable (Figure 1c(ii)). This configuration is foreseen to be the most favourable for SIBA trapping, since as the particle leaves the optical potential, the system crosses the resonance leading to an increase of photons coupled into the nanocavity. Remarkably, this regime has not been studied in previous plasmon trapping experiments.

## Results and Discussion

To experimentally reproduce these regimes, we used the confocal scans presented in Figure.1b and selected antennae 8, 10 and 12, which correspond to the blue-shifted, resonant and red-shifted regimes, respectively. Figure 1d shows an experimental time trace with a trapping event for each of these antennae. The black trace corresponds to the transmitted signal for an empty trap and in orange when a single GNP is trapped.

As expected from the earlier classification, the number of transmitted photons increases and decreases when the object is trapped under blue-shifted and red-shifted regimes, respectively. Similarly, the transmission oscillates around the empty trap value for the resonant regime. These results are in good agreement with the large frequency shifts computed above. In the following experiments, we focus our attention on the blue-shifted and resonant regimes, i.e. (i) and (ii), to determine which is better suited for trapping at low powers.

To characterize the optical potential we calibrated the stiffness of the system $\kappa_{tot}$. The standard procedure to calibrate the stiffness of an overdamped harmonic oscillator consists in recording the Brownian motion of the trapped object for few seconds and then fitting its power spectral density (PSD) to a Lorentzian curve [35]:

$$S_{PSD} = \frac{k_b T}{\gamma \pi^2 (f_c^2 + f^2)}$$

where $T$ is the temperature, $k_b$ the Boltzman constant and $f_c = \frac{\kappa_{tot}}{2\pi\gamma}$ the cutoff frequency [35]. In Figure 2a we compare the PSD of a 10s signal for a trapped particle (blue curve) with the one of an empty trap (grey curve) at 1.9mW. From the fit, we obtain $\kappa$ =4.51fN/nm, which corresponds to a normalized value of 2.4 fN/nm for an optical intensity of 1 mW/μm². Due to the large polarizability of our object, this normalized stiffness is the largest experimental value reported so far for plasmon trapping [24]. Although this characterization approach provides a stiffness value for the trap, it averages out any dynamic SIBA contribution due to the exceedingly large acquisition time (~s) compared to the trap relaxation time, which is $\tau = 1/f_c$. Therefore, a method to characterize the trap at shorter timescales is required.

To observe the reconfigurable nature of the trap, we applied the following binning procedure to our data: First we chose a bin time of 80 ms, which is two orders of magnitude above the trap relaxation time. Then, we computed, the autocorrelation function for each bin and linearly fit its logarithm to obtain the relaxation time as previously described [24]. Note that the PSD and the autocorrelation function form a Fourier pair, containing the same information. However, the former becomes simpler to fit due to its exponential decay. Figure 2b shows a sample of processed data for a ≈ 3s time trace. We distinguished two different groups of autocorrelation curves: in grey those corresponding to an empty trap and in blue those for a single trapped GNP. The linear fits are plotted in orange.

By removing the non-trapping events, we used the remaining values of $\tau$ to build a stiffness probability density function $\rho(\kappa_{tot})$ that contains the information about the modulation of the optical potential in presence of the trapped particle. Figure 3 shows $\rho(\kappa_{tot})$ at three different optical intensities for the blue-shifted and resonant regimes. The experimental distributions in both regimes (orange points) are perfectly fit by a normalized sum (black line) of two Lognormal distributions (blue and red), revealing the two different values of $\kappa_{tot}$. A radically different behavior is observed between these two regimes. In the blue-shift regime, $\rho(\kappa_{tot})$ is dominated by the red peak at high intensities (>0.6mW/μm$^2$). In this situation, the particle is highly confined (could be trapped for hours), thus no significant modulation occurs ($\kappa_{tot} \approx \kappa_{opt}$). When the incident intensity decreases (≈0.48mW/μm$^2$), $\kappa_{opt}$ becomes weaker allowing the particle to explore a wider region of the potential away from the equilibrium position. As a result, the overlap of the particle and the cavity mode decreases, blue-shifting the system resonance away from the excitation laser. This further decreases the stiffness of the optical potential and a new peak (blue), corresponding to the modulated potential, appears at lower $\kappa_{tot}$ values than the red one ($\kappa_{tot} = \kappa_{opt} + \kappa_{SIBA} < \kappa_{opt}$) (Figure 3a). Note that $\kappa_{SIBA}$ is negative, since when the particle leaves the antenna, the number of photons in the cavity decreases (c.f. Figure1c(i)). Finally at low powers, the GNP spends most of the time away from the trap equilibrium position as shown by the dominance of the blue peak, which suggests that the particle is nearly free diffusing and weakly trapped. This agrees with the fact that trapping events last very short times, typically <1s, as seen in Figure 2b. Figure 3a insets illustrate the potential seen by the GNP in the blue-shifted regime. For small displacements from the trap centre (red area) it experiences the restoring constant $\kappa_{tot} \approx \kappa_{opt}$. As it moves further away, the lower restoring constant $\kappa_{tot} = \kappa_{opt} + \kappa_{SIBA}$ reduces the slope of the potential (blue area).

In the resonant regime, the red peak also dominates at high intensities, where barely no modulation occurs ($\kappa_{tot} \approx \kappa_{opt}$). However, in this regime, when the laser intensity is lowered and the particle explores a larger region of the nanocavity, the system's resonance blue-shifts towards the laser line. Consequently, more photons couple into the nanocavity, modulating the potential ($\kappa_{tot} = \kappa_{opt} + \kappa_{SIBA}$) and increasing the optical forces that pull back the particle to the center of the trap. This is demonstrated by the fact that in Figure 3b the new peak appears at higher $\kappa_{tot}$ values ($\kappa_{tot} = \kappa_{opt} + \kappa_{SIBA} > \kappa_{opt}$) than the previous peak in Figure 3a. Finally, at very low powers (≈0.26mW/μm$^2$), the blue peak dominates and broadens to higher $\kappa_{tot}$ due to the larger modulation of the potential. This corresponds to a significant increase in the optical restoring forces, resulting in a more stable trap than the blue-shifted regime under the same intensities. Figure 3b insets illustrate the potential seen by the GNP in the resonant regime. For small displacements from the trap centre (red area) it experiences the restoring constant $\kappa_{tot} \approx \kappa_{opt}$. As it moves further away, the restoring constant $\kappa_{tot} = \kappa_{opt} + \kappa_{SIBA}$ increases the slope of the potential (blue area).

To further understand the dependence of the SIBA effect with the optical power, we plot in Figure 4 $\kappa_{tot}$ as a function of the optical intensity for both detuning regimes. In the blue-shifted regime (Figure 4a), we see that both $\kappa_{opt}$ and $\kappa_{opt} + \kappa_{SIBA}$ increase linearly with the optical excitation power, in agreement with previous observations in this regime [24]. In the resonant regime (Figure 4b) $\kappa_{opt}$ still grows linearly with the intensity, but now $\kappa_{opt} + \kappa_{SIBA}$ becomes inversely proportional to the intensity. This demonstrates that, as power is lowered, the SIBA effect becomes stronger until it becomes the main trapping mechanism, in agreement with the dominance of the blue peak at low powers in Figure 3b.

In other words, the relative trapping efficiency of each detuning regime is highly dependent of the optical intensity conditions. At high optical intensities the blue-shifted regime gives a stiffer trap. Conversely, the resonant regime becomes the most efficient as power decreases, reaching a stiffness up to 4 times higher.

## Conclusions

In conclusion, by enhancing the optomechanical interaction of a plasmonic nanocavity and the trapped object we have revealed the optomechanical origin of the SIBA effect. This allows us to direct observe the reconfigurable nature of the optical potential and to identify the optimum detuning regime that maximizes the trapping efficiency under low laser intensities. This results crucial for trapping and manipulation of objects extremely sensitive to photo-damage such as biological samples, fluorescent single emitters, etc. From the optomechanics perspective, the parameters of our system belong to a regime widely unexplored by the plethora of existing optomechanical systems[33], namely: an overdamped mechanical oscillator and a low quality factor optical nanocavity. Still, due to the subwavelength confinement of the optical field and the small size of the nanocavity, we achieve exceedingly large optomechanical coupling constants ($G \approx 750$GHz/nm) allowing highly efficient nanoparticle confinement and motion transduction. By properly engineering nanocavities with higher optical quality factors (narrower linewidths), optomechanical interactions can be enhanced, bringing further improvements in trapping performances at even lower optical powers.

## Acknowledgments

We acknowledge Lukas Neumeier and Darrick Chang for fruitful discussions and Raul Rica and Jaime Ortega for their comments on the paper. All authors acknowledge financial support from the Fundació Privada Cellex Barcelona, the Spanish Ministry of Economy and Competitiveness (grant FPU-AP-2012-3729 and FIS2013-46141-P) and the European Research Council through Consolidator grant n.64790.

# Figure 1

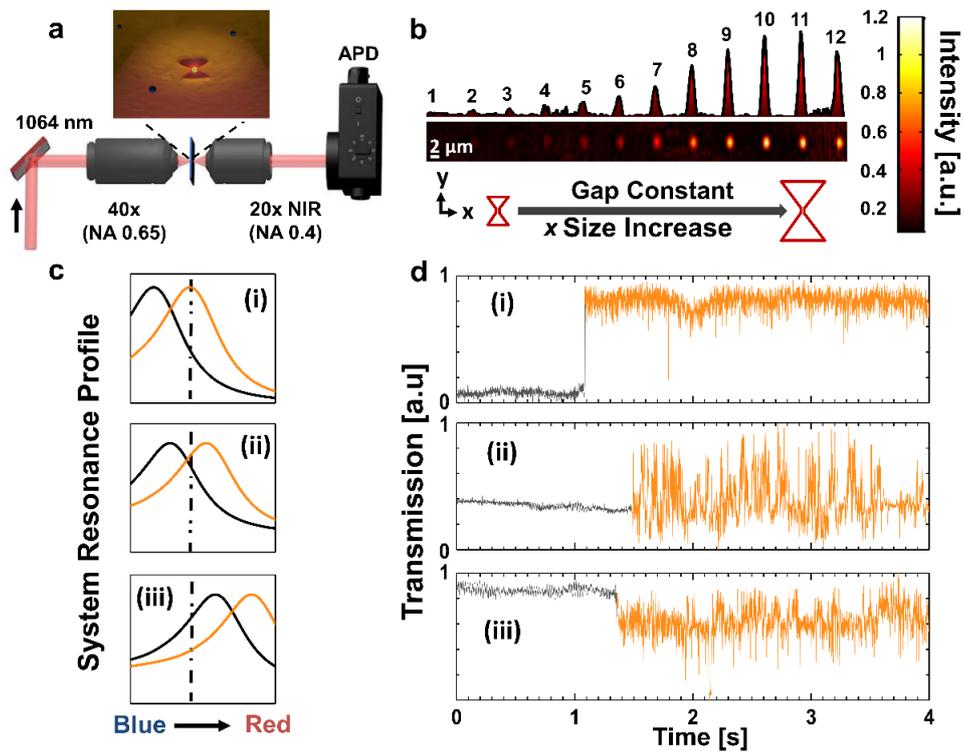

**Figure 1** Schematic view of the experimental configuration. (**a**) A 1064 nm linearly polarized laser is focused onto the sample at the antenna position with a 40x (0.65 NA) objective. The transmitted light is then collected with a 20x NIR (0.4 NA) objective and focused onto an avalanche photodiode (APD). (**b**) Experimental transmission map for different BNA with increasing size. The gap along the $x$ axis is fixed at 85 nm while the dimension of the antenna increases along the array. The polarization of the laser is aligned along $x$. (**c**) Cavity resonance shift for the three possible detuning regimes: (i) blue-shifted, (ii) resonant and (iii) red-shifted. The black trace corresponds to an empty trap and the orange one to a particle being trapped. The dashed line represents the excitation laser at 1064 nm. (**d**) Experimental transmission time traces for the three detuning regimes. We used the antennae labelled 8, 10 and 12 in (**b**).

**Figure 2**

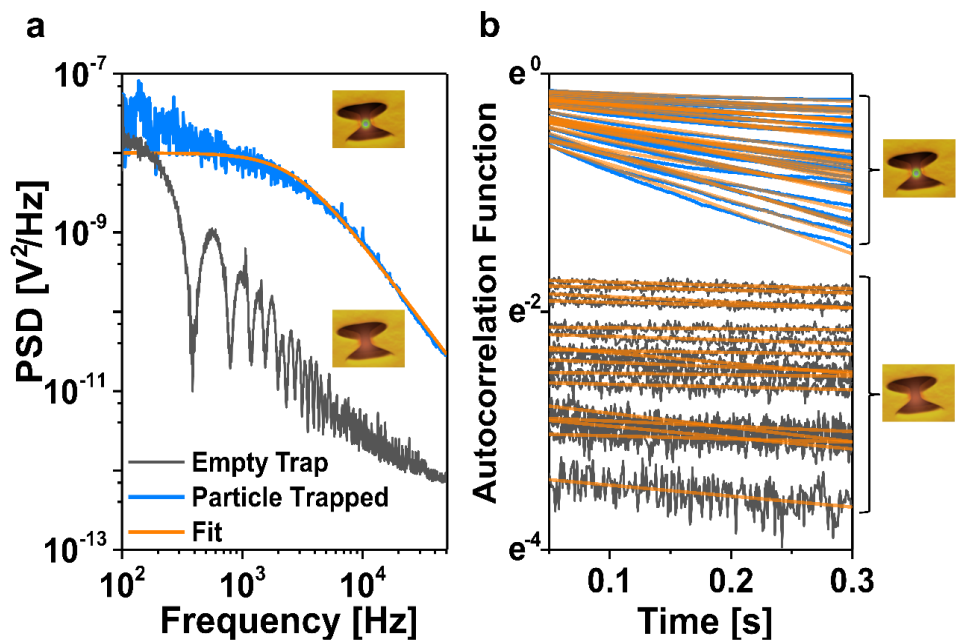

**Figure 2** Calibration of a plasmonic cavity trap. (**a**) Power spectral density for an empty trap (grey), and a single GNP trapped (blue). The orange line is a Lorentzian fit giving a trap stiffness of $\kappa$ =4.51 fN/nm. (**b**) Computed autocorrelation function of an 8s time trace of a trapped particle jumping in and out of a trap at 0.26 mW/μm². Blue lines correspond to trapping events, whereas empty trap events are shown in grey. The linear fits giving the relaxation time are plotted in orange.

# Figure 3

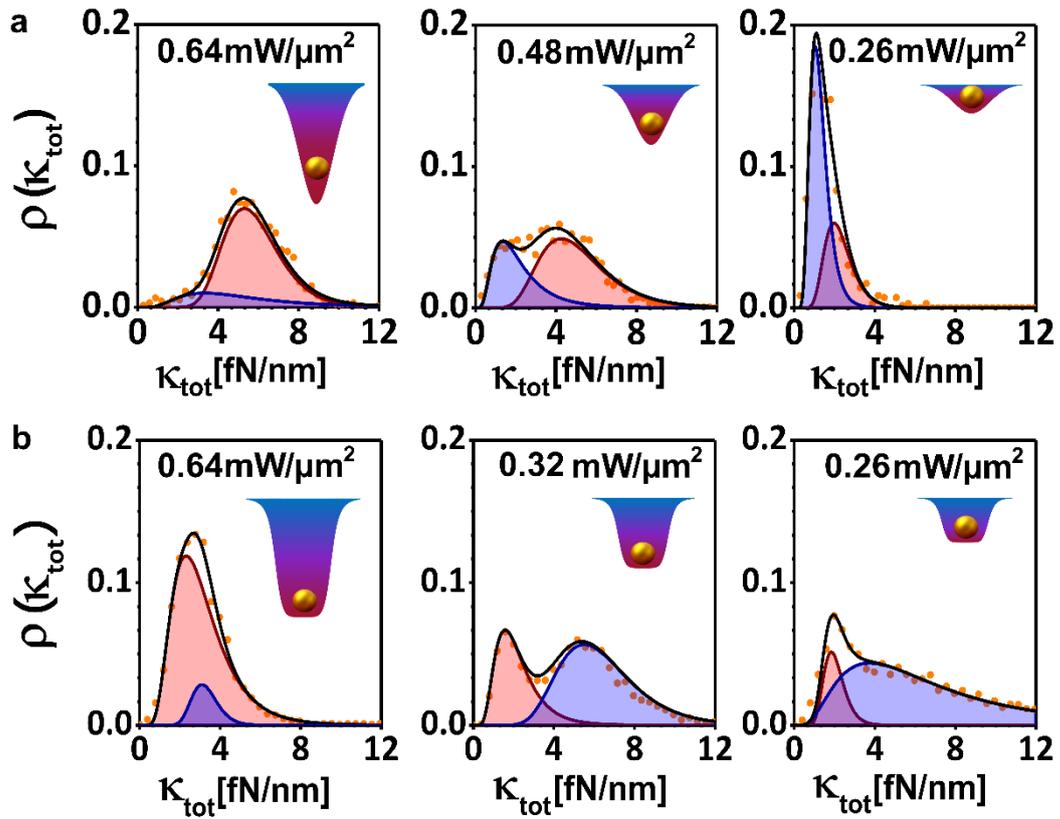

**Figure3** Probability distribution of the total stiffness $\kappa_{tot}$ at different powers. (**a**) Blue-shifted regime and (**b**) resonant regime. The experimental distributions (orange dots) are fitted by the sum of two Lognormal contributions (black line) at different optical intensities. The red peak represents the stiffness $\kappa_{opt}$ and the blue peak $\kappa_{opt} + \kappa_{SIBA}$. Each distribution is obtained using between 5000 (higher intensities) and 2000 (lowest intensities) fitted values of $\tau$. Insets show an impression of the GNP behavior in the modulated potential, where the blue (red) well correspond to the blue (red) peak contributions of $\kappa_{tot}$.

# Figure 4

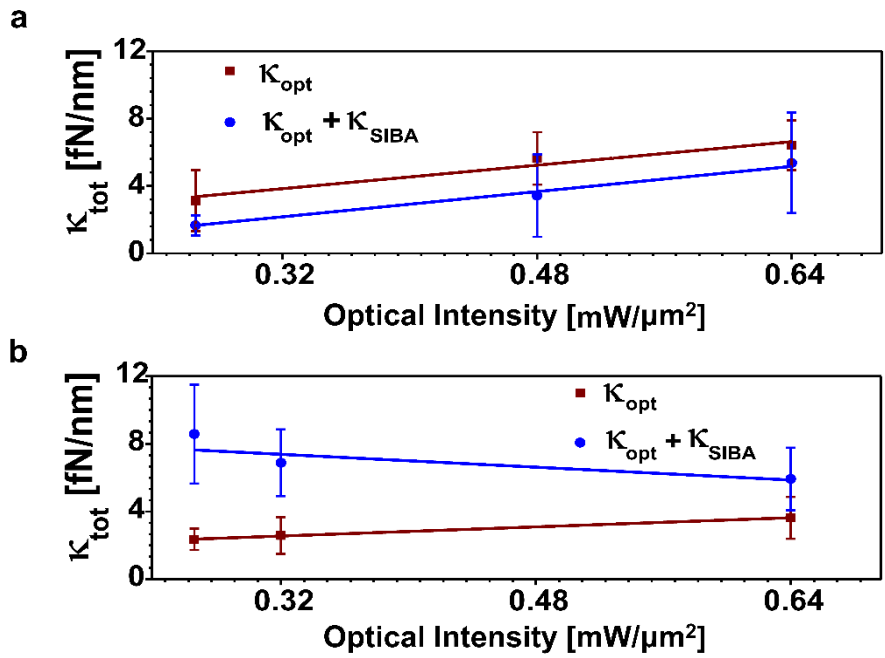

**Figure4** Stiffness as a function of the incident optical intensity. (**a**) For the blue shifted regime and (**b**) for the resonant regime. Error bars are computed from the standard deviation in the Lognormal distributions.